\documentclass[epj]{svjour}
\usepackage{graphics}
\begin{document}


%\catchline{}{}{}{}{}

%%%%%%%%%%%%%%%%%%%%%%%%%%%%%%%%%%%%%%%%%%%%%%%%%%%%%%%%%%%%%%%%%%%%

\title{ A Study of  Multi-$\Lambda$ Hypernuclei within Spherical Relativistic Mean-field Approach}
\author{Asloob A. Rather\inst{1}, 
M. Ikram\inst{1}\thanks{e-mail: {\texttt{ikram@iopb.res.in; ikramamu@gmail.com}}}
, A. A. Usmani\inst{1}, B. Kumar\inst{2,3}, S. K. Patra \inst{2,3}% etc
%%%%%%%%%%%\thanks{\emph{Present address:} Insert the address here if needed}%
}                     % Do not remove
%
%%%%%%%%%\offprints{}          % Insert a name or remove this line
%
\institute{Department of Physics, Aligarh Muslim University, Aligarh-202002, India. \and 
Institute of Physics, Bhubaneswar-751 005, India.
 \and Homi Bhabha National Institute, Anushakti Nagar, Mumbai - 400094, India.}
\date{Received: date / Revised version: date}

\abstract{
This research article is a follow up of earlier work by 
M. Ikram et al., reported in International Journal of Modern 
Physics E{ \bf{25}}, 1650103 (2016) wherein we searched 
for $\Lambda$ magic numbers in experimentally confirmed doubly magic 
nucleonic cores in light to heavy mass region 
(ie.$^{16}O - ^{208}Pb$) by injecting $\Lambda$'s into them. 
In present manuscript, working within the state-of-art relativistic mean 
field theory with inclusion of $\Lambda N$ and $\Lambda\Lambda$ interaction 
in addition to nucleon-meson NL$3^{*}$ effective force, 
we extend the search of lambda magic numbers in multi-$\Lambda$ 
hypernuclei using the predicted doubly magic nucleonic cores 
$^{292}$120, $^{304}$120, $^{360}$132, 
$^{370}$132, $^{336}$138, $^{396}$138 of elusive superheavy mass regime.
In analogy to well established signatures of magicity in 
conventional nuclear theory, the prediction of hypernuclear 
magicity are made on the basis of one-, two-$\Lambda$ separation 
energy ($S_\Lambda, S_{2\Lambda}$) and two lambda shell gaps 
($\delta_{2\Lambda}$) in multi-$\Lambda$ hypernuclei. 
The calculations suggest that the $\Lambda$ numbers 
92, 106, 126, 138, 184, 198, 240, and 258 might 
be the $\Lambda$ shell closures after introducing the 
$\Lambda$'s in elusive superheavy nucleonic cores. 
The appearance of new lambda shell closures other than 
the nucleonic ones predicted by various relativistic 
and non-relativistic theoretical investigations can be attributed to 
the relatively weak strength of spin-orbit coupling 
in hypernuclei compared to normal nuclei. 
Further, the predictions made in multi-$\Lambda$ hypernuclei under 
study resembles quite closely with the magic numbers in 
conventional nuclear theory suggested by various relativistic and 
non-relativistic theoretical models. 
Moreover, in support of $\Lambda$ shell closure the investigation of $\Lambda$ 
pairing energy and effective $\Lambda$ pairing gap has been made. 
We noticed a very close agreement of the predicted $\Lambda$ 
shell closures with the survey made on the
pretext of $S_{\Lambda}$, $S_{2\Lambda}$ and $\delta_{2 \Lambda}$
except for the appearance of magic numbers corresponding to
$\Lambda$ = 156 which manifest in $\Lambda$ effective pairing
gap and pairing energy. 
Also, lambda single-particle spectrum is analyzed to mark the 
energy shell gap for further strengthening the predictions made 
on the basis of separation energies and shell gaps.
Lambda and nucleon spin-orbit interactions are analyzed to confirm 
the reduction in magnitude of $\Lambda$ spin-orbit interaction
compared to the nucleonic case, however interaction profile is 
similar in both the cases. 
Lambda and nucleon density distributions have been investigated 
to reveal the impurity effect of $\Lambda$ hyperons
which make the depression of central density of the 
core of superheavy doubly magic nuclei.
Lambda skin structure is also seen.
\keywords{Hypernuclei, magic number, separation energy, spin-orbit interaction, 
relativistic mean field theory}
%\PACS{
%      {PACS-key}{discribing text of that key}   \and
%      {PACS-key}{discribing text of that key}
%     } % end of PACS codes
    } %end of abstract

\maketitle
\section{Introduction}
Hypernuclear physics has become an exciting area of research due to the 
epoch-making strides in the experimental arena. The primary goal of
strangeness nuclear physics is to understand the baryon-baryon interaction which
is fundamental and is very important for nuclear physics studies. The two-body scattering 
are quite useful for elucidating the nature of baryon-baryon interactions. Pursuing
this line of thought many NN scattering experiments have been performed and the total 
number of NN scattering are more than 4000. However, because of the difficulty
encountered in carrying out the nucleon-hyperon and hyperon-hyperon scattering experiments, 
the YN data available is very scarce. To be more precise, the number of differential
cross-section data is nearly about 40 and there is no YY scattering data.
Further, the YN and YY potential models put forth so far have large ambiguities
in their predictions. Thus, it is imperative and wise to study the structure of 
single-$\Lambda$, double-$\Lambda$ and multi-$\Lambda$ hypernuclei rather than constraining 
the various theoretical models employing YN and YY potentials.
As the number of data points in hyperon-nucleon sector is few, it is 
quite difficult to extract the information regrading scattering lengths even. 
This necessitated to analyze this small data using symmetry of SU(3)$_f$ models 
of baryon-baryon interaction which establishes close connections with these data in NN sector. 
However, realistic models must include the symmetry breaking terms
as it is badly broken owing to the difference in mass between s and (u,d) quark. 
Keeping in view above consideration, various YN potential models have been developed 
along these lines for hypernuclear studies. 
The well known models are, the Nijmegen models including the hard core models 
D~\cite{NRS77} and F~\cite{NRS79}, the soft core models NSC89~\cite{MRS89}, 
NSC97~\cite{RSY99} and the extension of soft 
core models like ESC04~\cite{RY06a}, ESC08~\cite{NRY15b} which consider two-meson 
exchanges and other short range contributions in addition to one-boson exchange. 
In particular, these models enable extension to hyperon-hyperon potentials, where 
there is no scattering data available~\cite{NRY15a}, thus implying an 
unfortunate increase in model dependence. 
The other type of models known by Bonn-Julich
multi-meson exchange models~\cite{HHS89,RHS94,HM05} which are based 
on SU(6) symmetry of quark-model. 
However, the short range nature of YN interaction in these models and Nijmegen 
models are governed by the way scalar meson interactions are introduced and 
thus imperactively has got model dependence. 
The Effective theory chiral model of Leading Order LO~\cite{PHM06} and Next 
Leading Order NLO~\cite{HPKMNW13} are another class of model used for 
hypernuclear studies that makes use of regularized PS Goldstone-boson exchange YN 
potentials and adding zero-range contact terms for the parameterization of YN 
coupled-channel interactions and details are reported in Ref.~\cite{H13}.
Further, a quark-model baryon-baryon potential which respects 
SU(6) symmetry~\cite{FSN07} can be used for the construction of 
hyperon-nucleus potentials~\cite{KF09}.

In nuclear physics, the nuclei with magic numbers have been a hot topic
since the birth of this subject~\cite{M49,HJS49}. 
With the advent of state-of-art facility
radioactive nuclear ion beams (RNIBs) the quenching of traditional magic
numbers and appearance of new magic numbers have been observed in nuclei
with exotic isospin ratios. 
%The study of nuclei with magic numbers
%in neutron rich region have got huge implication for the study of
%astrophysical r-process~\cite{BBFH57}.
The explanation of unusual stability attributed to nuclei by neutron(proton)
numbers: 2, 8, 20, 28, 50, 82 and 126 commonly referred to as magic numbers
in literature, was made in non-relativistic shell model
by 3D harmonic oscillator central potential in conjunction
with a very strong spin orbit interaction added manually~\cite{M49,HJS49,BM69,PS80,H99}. 
Contrary to this, the models which come under the aegis of 
relativistic framework say the relativistic mean-field (RMF)
~\cite{W74,SW86,R89,HTWTC91,R96} in which the strong spin-orbit 
interaction is a natural outcome of the interplay
between strong scalar and vector potentials, which are indispensable
for reproducing the saturation properties of matter. The shell structure
is obtained due to proper setting of scalar and vector potentials
without incorporating any additional parameter for spin-orbit splitting.
The precise location of shell and subshell closures in nuclear landscape has provided 
significant inputs needed for the development of nuclear structure models and theory. 
However, the position of shell closures is not static particularly for exotic nuclei 
where there are large number of neutrons relative to protons. 
The ongoing experimental endeavours are striving for the location of these shell 
closures in previously unknown nuclei and trace the movement of single-particle 
orbitals which are responsible for this dynamic shell structure. 
However, we are lucky enough to have a number of experimental observables that 
show variation at shell closures across the nuclear chart. 
The nucleon magicity can be inferred from the discontinuity in these 
observables like separation energy (one- and two-neutron separation energies), 
two nucleon shell gap, the energies of $\alpha$ and $\beta$ transitions 
(nuclear ground-state moments), shell correction energy, pairing correction, 
effective pairing gap, binding energy and excitation energy of 
low-lying vibrational states. 
Further, nuclear quadrupole moment provides a measure of nuclear charge 
distribution and hence the nuclear shape. 
The ground-state quadrupole moment goes to zero at the shell closures 
and thus shape of nuclei is expected to be spherical; however 
far away from shell closures the nuclei may be axially deformed 
and might possess large value of quadrupole moment. 
The early theoretical investigation employing macro-microscopic 
approach predicted the spherically doubly magic nucleus $^{298}$114. 
However, the existence analysis of magicity in nuclear landscape 
using different models revealed that showed volatility (variation) in 
the prediction of shell closures and many reports of investigation 
were successful in establishing the fact that nuclear magicity 
is dependent on both model and effective force employed. 
The inference that can be drawn from these studies that reliability 
of different parameterizations of the models and the accuracy with 
which the single-particle energies are estimated plays an indispensable 
role in heavy and superheavy nuclei owing to the large level density in such nuclei. 
Thus, it can be concluded that magic numbers in nuclear landscape are 
fragile/localized and search-space dependent implying that different search space 
would possibly result in the emergence of different magic numbers.

There is wealth of literature available  regarding the theoretical 
attempts being made for identifying the nuclear magic numbers in superheavy region. 
Some are as follows: 
Within the parameterizing nucleon mean field in the form of Woods-Saxon 
potential Brack et. al.~\cite{BGH85} explored the nuclear magicity in 
a wide range of nuclear chart and suggested the proton magicity in the 
region of Z = 114, 120 and neutron magicity at N = 184. 
In the same vein, the proton magic number at Z = 114 and neutron 
magic number at N = 184 was confirmed in the 
works of Refs.~\cite{S94,MNMS95,MNK97,D05}. 
The microscopic self-consistent calculations in Hartree-Fock-Bogoliubov 
(HFB) with effective Gogny forces yielded proton magic numbers at
Z = 114, 120 and 126 and N = 164, 184, 228~\cite{BBGD01}. 
%To be more explicit, SKM predicted magicity at Z = 120; while SKI1, SKI3 
%and SKI4 yielded proton magicity at Z = 120. 
The parameterizations SLy4 and SkP resulted the proton magicity at Z = 124 
and 126 while SkO predicted at Z = 126~\cite{BNR01,BRRMG99,BRRMG98}. 
Further, all these parameterization driven calculations
predicted neutron magic number at N = 184~\cite{BNR01,BRRMG99,BRRMG98,KBNRVC00}.
%By performing calculations within the framework of semi-empirical shell model 
Liran et. al.~\cite{LMZ00,LMZ000} predicted proton magicity at 
Z = 126 and neutron magicity at N = 184 within the semi-empirical shell model. 
Calculations within the relativistic HFB approach with finite range pairing force of 
Gogny effective interaction D$_1$ and effective interaction NLSH~\cite{LSRG96} 
established that Z = 114, and N = 160, 166, 184 combinations exhibit high stability 
compared to their neighbours and further doubly magic character at 
Z = 106 and N = 160 was predicted in Ref.~\cite{LSRG96}.
The non-relativistic skyrme Hartree Fock (SHF) with effective forces SkP and 
SLy7 predicted the nuclear magic number at Z = 126 and N = 184 and in addition 
presented the evidence of increased stability around N = 162 owing to 
the deformed shell effects~\cite{CDMNH96}.
%Calculations done for two-nucleon gaps within Skyrme-Hartree Fock 
%approach with SkM$^{*}$, SkP, SLy6, SkI1, SkI3, SkI4 effective interactions 
%and relativistic mean-field calculations performed using PL-40, NLSH, NL-Z, 
%TM1 effective interaction, 
K. Rutz et. al.~\cite{RBBSRMG96}, made a prediction about the doubly magic 
character of $^{298}$114, $^{292}$120 and $^{310}$126. 
Several other theoretical attempts employing relativistic mean-field 
models~\cite{BNR01,BRRMG99,BRB98,KBN00} yielded the magic numbers 
at Z = 120 and N = 172, 184 while the relativistic model in Ref.~\cite{SPSCV04} 
predicted the nuclear magicity at Z = 114, 120 and N = 172, 184, and 258 and the 
reconfirmation of closed shells at Z = 120 and N = 172, 184, 258 are given in Ref.\cite{mbhuyan}. 
The exhaustive investigations performed within the framework of continuum 
relativistic Bogoliubov theory~\cite{ZMZGT05} using various effective 
forces predicted proton numbers Z = 106, 114, 120, 126, 132, and 138 
and neutron numbers N = 138, 164, 172, 184, 198, 216, 228, 238, 252, 258 and 
274 as possible magic numbers. 
The survey made by Denisov on the basis of shell correction
calculations ~\cite{D05} predicted spherical proton magic numbers at 
Z = 82, 114, 164, 210, 274, 354 and spherical neutron magic numbers at 
N = 126, 184, 228, 308, 406, 524, 644, 772. 
Moreover, an extensive study by Nakada and Sugiura ~\cite{NS14,N13}
over a wide range of even-even nuclei within spherical
relativistic mean field approaches with semi-realistic interactions
(M3Y-P6 and P7 parameter sets)has been made and they choose the 
proton(neutron) pairing energy correction and suggested the possible 
magicity features in the region Z = 14, 16, 34, 38, 40, 58, 
46, 92, 120, 124, 126 and N = 40, 56, 90, 124, 172, 178, 164, 184 which 
show remarkable agreements with the results in Refs.~\cite{BGG91,GHGP09} 
where Gogny DIS and DIM interactions was used respectively.
Further, Ismail et.al.,~\cite{IEAA16} have also made more exhaustive 
investigation about to search of nucleonic closed shell within ultra 
heavy region and predicted a huge number of magic candidates. 
These predictions have been made on the basis of separation energy, shell 
gaps, pairing energy and shell correction energy etc.
%It would therefore be relevant to extend the thought of magicity in strangeness sector.
It has to mention that in previous work~\cite{ikram2016} we have 
inferred lambda magicity is in quite close agreement with nucleon 
magicity which arouse us to investigate lambda magic number 
in more exotic region and its predictions would give the strong 
evidence about the confirmation of nuclear magicity in superheavy regime.
The work in present manuscript is also suppose to be another 
attempt in support of nuclear magicity of superheavy region.  
Among the different methodologies used for identifying the nuclear magicity,
we adopt the calculations of one-, two- $\Lambda$ separation energies, two $\Lambda$
shell gaps, lambda pairing energy and effective lambda pairing gap to study
lambda magicity for the considered set of multi-$\Lambda$ hypernuclei. 
Thus, in this paper, our main emphasis would be to make extensive investigation 
to search the $\Lambda$ magic number in multi-$\Lambda$ hypernuclei 
within the spherical relativistic mean field approach and to suggest 
the confirmation of nucleon magic number of superheavy region within 
the favour of $\Lambda$ magicity. 
The  subject matter of manuscript is organised as follows: 
The theoretical framework of spherical relativistic mean field model 
is outlined in sect. two. 
Sect. three contains the results and discussion. 
Finally, the paper is summarized in sect. four.

%%%%%%%%%%%%%%%%%%%%%%%%%%%%%%%
\section{Theoretical formalism}
Relativistic mean field theory has established itself as a promising 
theoretical framework to study the infinite nuclear systems, finite 
nuclei including the superheavy mass region, hypernuclei and multi-strange 
systems~\cite{sugahara1994,vretenar1998,lu2003,win2008,ikram14,glendenning1993,mares1994,rufa1990}.
The suitable extension to strangeness sector for studying the hypernuclei 
and multi-$\Lambda$ hypernuclei is accomplished by incorporating the lambda-baryon 
as well as lambda-lambda interactions Lagrangian with effective $\Lambda$N 
and $\Lambda$$\Lambda$ potentials in addition to nucleon-meson NL3* force 
parameter and such type of attempts have already been 
made~\cite{sugahara1994,vretenar1998,lu2003,win2008,ikram14,glendenning1993,mares1994,rufa1990}. 
For quantitative description of the multi-$\Lambda$ hypernuclei additional 
strange scalar ($\sigma^*$) and vector ($\phi$) mesons have been incorporated
in the the Lagrangian which simulate the $\Lambda \Lambda$ interaction
~\cite{schaffner2002,shen2006,schaffner1994,schaffner1993}. 
Thus, the total Lagrangian density can be expressed as
{\footnotesize
\begin{eqnarray} 
\mathcal{L}&=&\mathcal{L}_N+\mathcal{L}_\Lambda+\mathcal{L}_{\Lambda\Lambda} \;, 
\end{eqnarray}
\begin{eqnarray}
{\cal L}_N&=&\bar{\psi_{i}}\{i\gamma^{\mu}
\partial_{\mu}-M\}\psi_{i}
+{\frac12}(\partial^{\mu}\sigma\partial_{\mu}\sigma
-m_{\sigma}^{2}\sigma^{2})		   
-{\frac13}g_{2}\sigma^{3}                  \nonumber \\
&&-{\frac14}g_{3}\sigma^{4}
-g_{s}\bar{\psi_{i}}\psi_{i}\sigma 		
-{\frac14}\Omega^{\mu\nu}\Omega_{\mu\nu}
+{\frac12}m_{\omega}^{2}\omega^{\mu}\omega_{\mu}		\nonumber \\
&&-g_{\omega }\bar\psi_{i}\gamma^{\mu}\psi_{i}\omega_{\mu}    
-{\frac14}B^{\mu\nu}B_{\mu\nu} 
+{\frac12}m_{\rho}^{2}{\vec{\rho}^{\mu}}{\vec{\rho}_{\mu}}  
-{\frac14}F^{\mu\nu}F_{\mu\nu}                        \nonumber \\
&&-g_{\rho}\bar\psi_{i}\gamma^{\mu}\vec{\tau}\psi_{i}\vec{\rho^{\mu}} 
-e\bar\psi_{i}\gamma^{\mu}\frac{\left(1-\tau_{3i}\right)}{2}\psi_{i}A_{\mu}\;, \\ 
\mathcal{L}_{\Lambda}&=&\bar\psi_\Lambda\{i\gamma^\mu\partial_\mu-m_\Lambda\}\psi_\Lambda
-g_{\sigma\Lambda}\bar\psi_\Lambda\psi_\Lambda\sigma  
-g_{\omega\Lambda}\bar\psi_\Lambda\gamma^{\mu}\psi_\Lambda \omega_\mu \;  \\
\mathcal{L}_{\Lambda\Lambda}&=&{\frac12}(\partial^{\mu}\sigma^*\partial_{\mu}\sigma^*  
-m_{\sigma^*}^{2}\sigma^{*{2}})
-{\frac14}S^{\mu\nu}S_{\mu\nu}
+{\frac12}m_{\phi}^{2}\phi^{\mu}\phi_{\mu}    \nonumber \\
&&-g_{\sigma^*\Lambda}\bar\psi_\Lambda\psi_\Lambda\sigma^*
-g_{\phi\Lambda}\bar\psi_\Lambda\gamma^\mu\psi_\Lambda\phi_\mu \;,      
\end{eqnarray}}
where $\psi$ and $\psi_\Lambda$ denote the Dirac spinors for 
nucleon and $\Lambda$ hyperon, whose masses are M and
$m_\Lambda$, respectively.
Due to the isoscalar nature of $\Lambda$ hyperon, it does not couple to ${\rho}$- mesons.
The quantities $m_{\sigma}$, $m_{\omega}$, $m_{\rho}$, $m_{\sigma^*}$, 
$m_{\phi}$ represent the masses of $\sigma$, $\omega$, $\rho$, $\sigma^*$, $\phi$
 mesons and $g_s$, $g_{\omega}$, $g_{\rho}$, $g_{\sigma\Lambda}$,
 $g_{\omega\Lambda}$, $g_{\sigma^*\Lambda}$, 
$g_{\phi\Lambda}$ are their coupling constants, respectively. 
The nonlinear self-coupling of ${\sigma}$-mesons is designated by 
by $g_2$ and $g_3$. 
The total energy of the system is given by 
%\begin{eqnarray}
$E_{total} = E_{part}(N,\Lambda)+E_{\sigma}+E_{\omega}+E_{\rho}
+E_{\sigma^*}+E_{\phi}+E_{c}+E_{pair}+E_{c.m.},$
%\end{eqnarray}
where $E_{part}(N,\Lambda)$ is the sum of the single-particle
 energies of the nucleons (N) and hyperons ($\Lambda$).
The energies parts $E_{\sigma}$, $E_{\omega}$, $E_{\rho}$, $E_{\sigma^*}$, 
$E_{\phi}$, $E_{c}$, $E_{pair}$ and $E_{cm}$ are the 
contributions of meson fields,Coulomb field, pairing energy and the
center of mass energy, respectively. There is wealth of available literature
about RMF parameterizations for predicting the nuclear ground state properties.
In the present manuscript, 
NL3* parameterization is employed for meson-baryon
coupling constant throughout the calculations~\cite{lalazissis09}. 
To find the numerical values of used $\Lambda-$meson coupling constants, 
we adopt the nucleon coupling to hyperon couplings ratio 
defined as; $R_\sigma=g_{\sigma\Lambda}/g_\sigma$, $R_\omega=g_{\omega\Lambda}/g_\omega$, 
$R_{\sigma^*}=g_{\sigma^*\Lambda}/g_\sigma$ and $R_\phi=g_{\phi\Lambda}/g_\omega$.
The relative coupling values are used as $R_\omega=2/3$, $R_\sigma=0.6104$, 
$R_\phi=-\sqrt{2}/3$ and $R_{\sigma^*}=0.69$~\cite{schaffner1994,dover1984,chiapparini09}. 
%The coupling constants of the hyperons to vector mesons
%is fixed using SU(6) symmetry under naive quark model implying that the vector coupling constants
%scale with the number of light quarks in the baryon in the form given above. 
The coupling constants of hyperons to vector mesons have to be 
compatible with the maximum neutron star masses and  are 
fitted to the $\Lambda$ binding energy in nuclear matter~\cite{GM91}. 
In present calculations, we use the constant gap BCS approximation to 
include the pairing interaction and the centre of mass 
correction is included by  $E_{cm}=-(3/4)41A^{-1/3}$.

%%%%%%%%%%%%%%%%%%%%%%%%%%%%%%%
\section{Results and discussions}
In present manuscript, we made an attempt to look for magic 
candidates in multi-$\Lambda$ hypernuclei under inspection 
on the basis of one-, two-$\Lambda$ separation energies, two 
lambda shell gaps and pairing energy/gaps etc 
by performing calculations within the self-consistent relativistic 
mean-field model with NL3$^{*}$ parameterization with inclusion 
of $\Lambda$N and $\Lambda$$\Lambda$ interactions. 
Our quest of hunt for magic numbers in multi-$\Lambda$ hypernuclei with 
doubly magic superheavy nucleonic cores receives the motivation from the 
seminal work of Zhang et. al.,~\cite{ZMZGT05} wherein they made an extensive 
investigation by performing calculations within relativistic continuum 
Hartree-Bogoliubov theory and established 
that Z = 120, 132, 138 and N = 172, 184, 198, 228, 238 and 258 might 
be the possible nucleonic magic numbers.
Further, Ismail et.al.,~\cite{IEAA16} have made more exhaustive investigation about 
to search of nucleonic closed shell within ultra heavy region and 
predicted a huge number of magic candidates of ultra heavy region. 
We extend the thought of nuclear magicity towards strangeness sector with 
more exotic lambda region and try to predict a triple magic multi-$\Lambda$ system 
having doubly magic superheavy nucleonic core.

%%%%%%%%%%%%%%%%%%%%%%%%%%%%%%%%%%%%%%%%%%%%%%%%%%%%%%%%%%%55
\subsection{Lambda separation energy and two lambda shell gaps}
\begin{figure}
\vspace{0.75cm}
\resizebox{0.50\textwidth}{!}{%
  \includegraphics{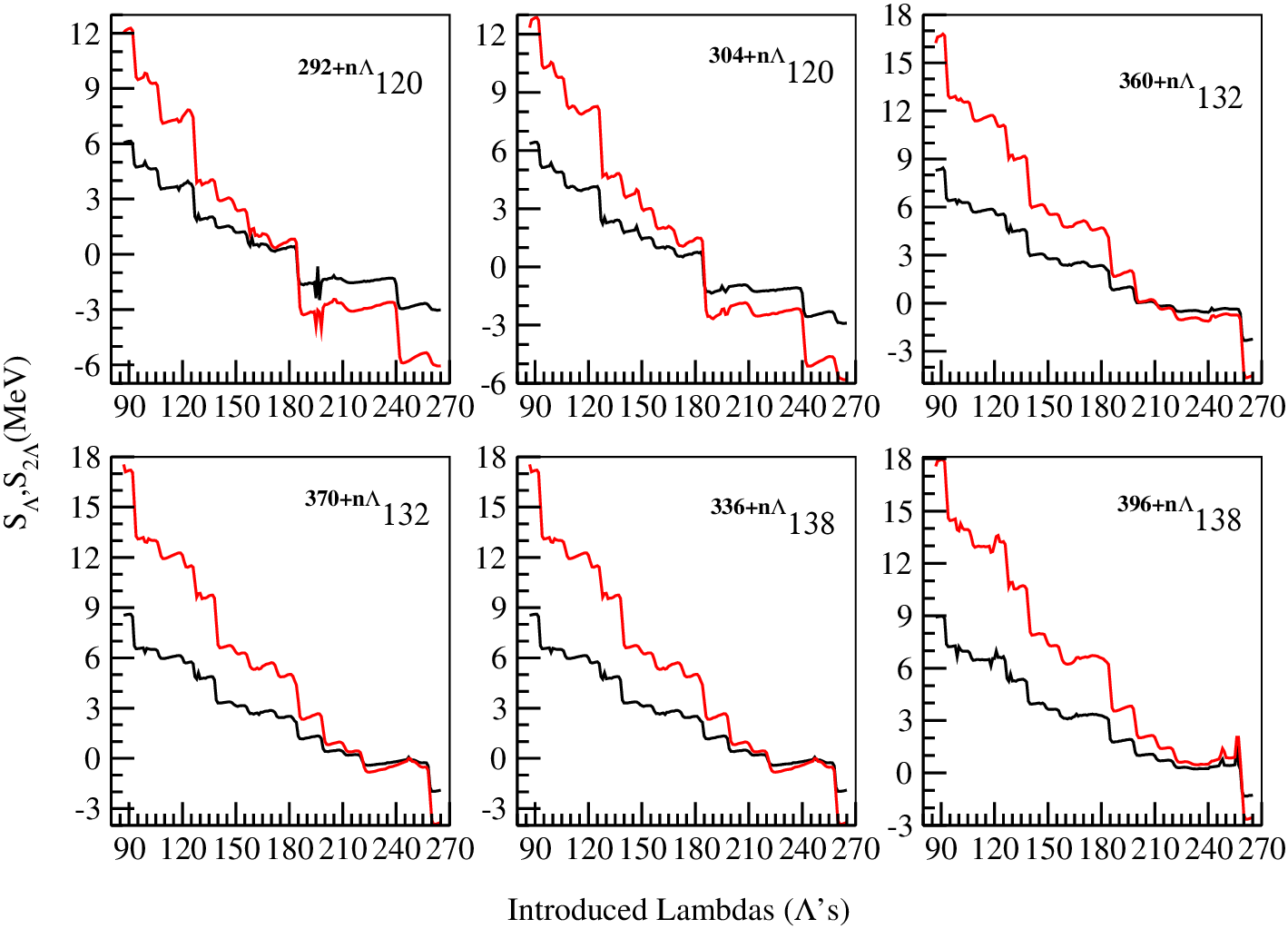}
}
\caption{(color online) One- and two-$\Lambda$ separation 
energy for multi-$\Lambda$ hypernuclei as a function of 
introduced $\Lambda$'s. Black lines represent the $S_\Lambda$ 
while $S_{2\Lambda}$ is shown by red lines.}
\label{sep}
\end{figure}

\begin{figure}
\vspace{0.75cm}
\resizebox{0.49\textwidth}{!}{%
  \includegraphics{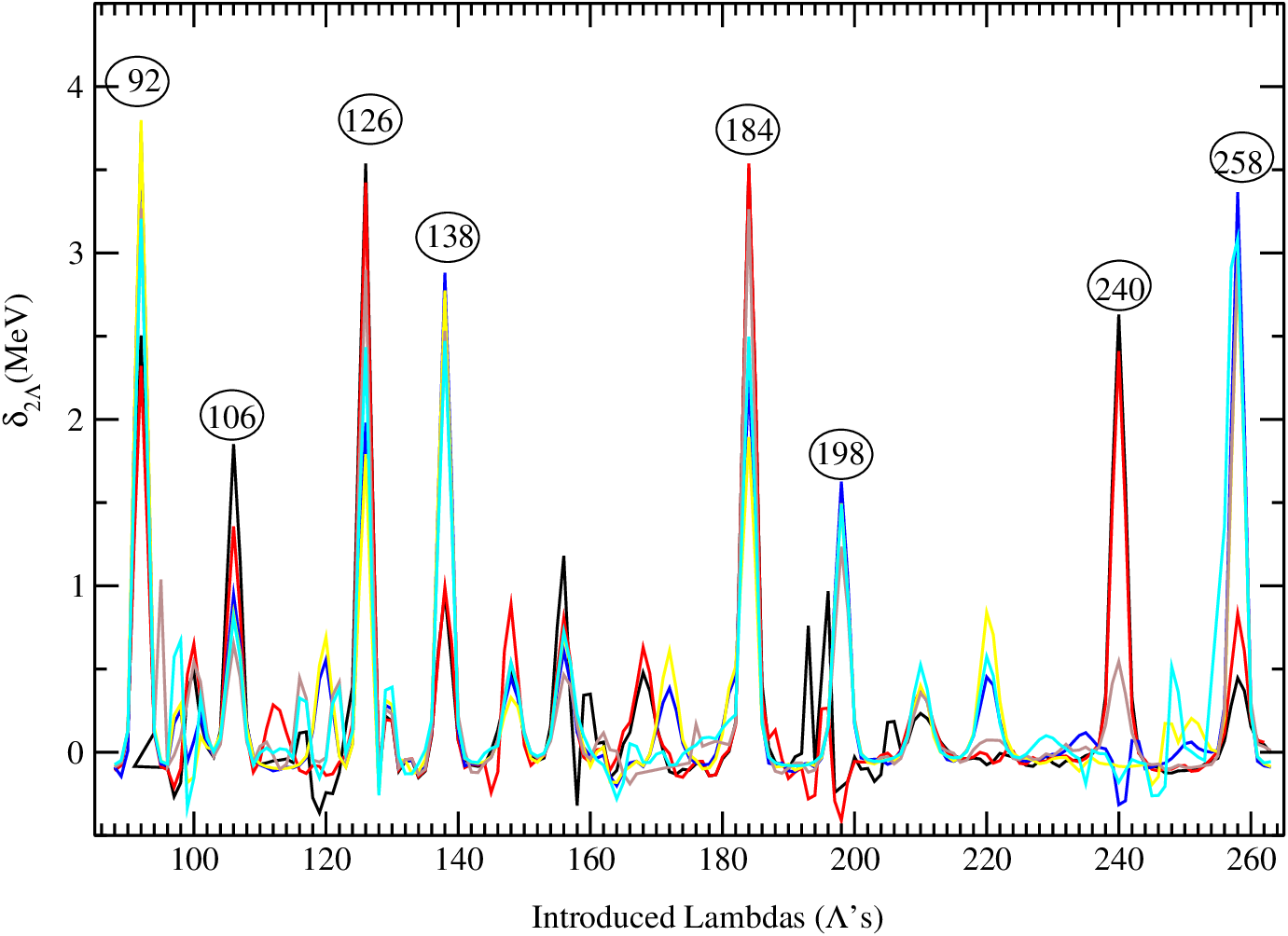}
}
\caption{(color online) Two-lambda shell gap for considered 
superheavy mass multi-$\Lambda$ 
hypernuclei as a function of introduced $\Lambda$'s.}
\label{delta}
\end{figure}

The separation energy is an important physical quantity and is of 
paramount importance for locating and identifying the magic numbers 
in nuclei as well as hypernuclei. 
The magic character is manifested  by the presence of large shell 
gaps in the single-particle spectrum of nuclei(hypernuclei). 
This in turn implies that the nucleons (hyperons) in the lower energy levels 
has got more binding energy than the nucleons(hyperons) in upper energy levels.
The reliable measures of one- and two-nucleon separation energies for 
marking the shell closures in conventional nuclear theory can be conveniently 
carried out to strangeness sector and may be expressed as follows.
{\footnotesize
\begin{equation}
S_{\Lambda}(N,Z,\Lambda) = BE(N,Z,\Lambda)- BE(N,Z,\Lambda-1)
\end{equation}
and,
\begin{equation}
S_{2\Lambda}(N,Z,\Lambda) = BE(N,Z,\Lambda)- BE(N,Z,\Lambda-2)
\end{equation}
}%end of footnotesize
These quantities are plotted in Fig~\ref{sep}. 
Moreover, two-$\Lambda$ separation energy is more efficient marker of 
magic numbers than one-$\Lambda$ separation 
energy due to the absence of odd-even staggering. 
For the considered lambda chain ($\Lambda$ = 90 - 265), 
the binding energy increases up to a certain limit by increasing 
the number of injected $\Lambda$'s in considered doubly magic nucleonic cores. 
However, for a fixed N, Z combination, the $S_{\Lambda}$ and $S_{2\Lambda}$ show a 
gradual decreasing trend with the number of injected $\Lambda$'s. 
A sudden fall in $S_{\Lambda}$ and $S_{2\Lambda}$ in analogy to 
neutron and proton chains indicates the occurrence of $\Lambda$ shell closure. 
The abrupt decrease in the value of $S_{2\Lambda}$ at $\Lambda$ 
= 92, 106, 126, 132, 138, 198, 228, 240, 258 can be easily seen in the 
multi-$\Lambda$ hypernuclei under investigation thus revealing their magic character. 
Thus, these numbers correspond to $\Lambda$ magic numbers in the 
multi-$\Lambda$ hypernuclei and form the triply magic 
system with doubly magic superheavy nucleonic cores.
In order to provide a clear presentation of magicity features, the 
shell effects are also quantified in terms of two lambda shell 
gaps which is a carry over from the conventional nuclear physics. 
The two lambda shell gap is expressed as
{\footnotesize
\begin{eqnarray}
\delta_{2\Lambda}(N,Z,\Lambda) &=& 2BE(N,Z,\Lambda)-BE( N,Z,\Lambda + 2)- BE(N,Z,\Lambda-2) \nonumber \\
                               &=& S_{2\Lambda}(N,Z,\Lambda) - S_{2\Lambda} (N,Z,\Lambda+2) 
\end{eqnarray}
}
The two lambda shell gap is considered to be strong signature for 
identifying the magic numbers. 
Two lambda shell versus number of injected lambdas is plotted in 
Fig.~\ref{delta} for all the considered multi-$\Lambda$ hypernuclei under investigation. 
A peak in the two lambda shell gap indicates the drastic change in 
two-$\Lambda$ separation energies. 
A peak at certain $\Lambda$ number hints towards the $\Lambda$ shell closure. 
However, the quality of $\Lambda$ shell gap is dictated by 
the sharpness and magnitude of the peak.
From figure~\ref{delta}, the peaks at $\Lambda$ = 92, 106, 126, 138, 184, 
198, 240 and 258 are evident and these numbers are suppose to be 
$\Lambda$ magic numbers.

%%%%%%%%%%%%%%%%%%%%%%%%%%%%%%%%%%%%%%%%%%%%%
\subsection{Lambda pairing energies and pairing gaps}
Pairing is an important ingredient in a mean-field description 
of nuclear structure calculations~\cite{BMP58}.
It is worthy to mention that the pairing correlation is 
significant for open shell nuclei with a large density of almost
degenerated states, which offers an opportunity to residual two-body interaction
to mix these states in order to produce a unique ground-state~\cite{RO84}.
This means the pairing contribution in total energy is very less  
or almost zero for closed shell nuclei.
The two mostly used methods to treat pairing are HFB and BCS approximation.
In HFB, pairing correlations are taken into account by 
introducing the concept of quasi-particles
defined by Bogoliubov transformation while as BCS approximation is just
a simplification of HFB for time-reversal invariant systems~\cite{BHR03}.
In both of these schemes, each single nucleon state $\phi_{\alpha}$
is associated with an occupation amplitude $v_{\alpha}\in[0,1]$
where the complementing non-occupation amplitude is $u_{\alpha} = \sqrt{1-v^{2}_{\alpha}}$.

This refer to that pairing energy may serve as a reliable source of 
information for identifying the closed shell nuclei or hypernuclei.
In BCS approximation, the $\Lambda$ pairing energy is defined in 
terms of effective pairing gap $\Delta$ and occupations as given by 
\begin{equation}
E_{pair} = - \Delta \sum_{\alpha} \sqrt{w_{\alpha}(1-w_{\alpha})}.
\end{equation}
In order to ensure successful calculation of every desired nucleus
occupation numbers w$_{\alpha}$ are allowed to vary between 0 and 1, 
whose values are determined using schematic pairing within the 
constant gap approach using single-particle energy spectrum~\cite{BF76}
\begin{equation}
w_{\alpha} = \Big(\frac{1}{2}+\frac{\epsilon_{\alpha}-\epsilon_{Fermi}}{\sqrt{(\epsilon_{\alpha}-\epsilon_{Fermi})^{2}+\Delta^{2}}}\Big)^{1/2}
\end{equation}
where 
\begin{equation}
\Delta = \frac{11.2 MeV}{\sqrt{A}}
\end{equation}
and the Fermi energy is determined by
\begin{equation}
\sum_{\alpha =1}^{\Omega}W_\alpha = \left\{
\begin{array}{rl}
number of protons or\\
number of neutrons
\end{array} \right.
\end{equation}
where $\Omega$ is the number of shell model states incorporated for
protons, or for neutrons respectively and further, we always
include one shell above the last closed shell.

\begin{figure}
\vspace{0.75cm}
\resizebox{0.50\textwidth}{!}{%
  \includegraphics{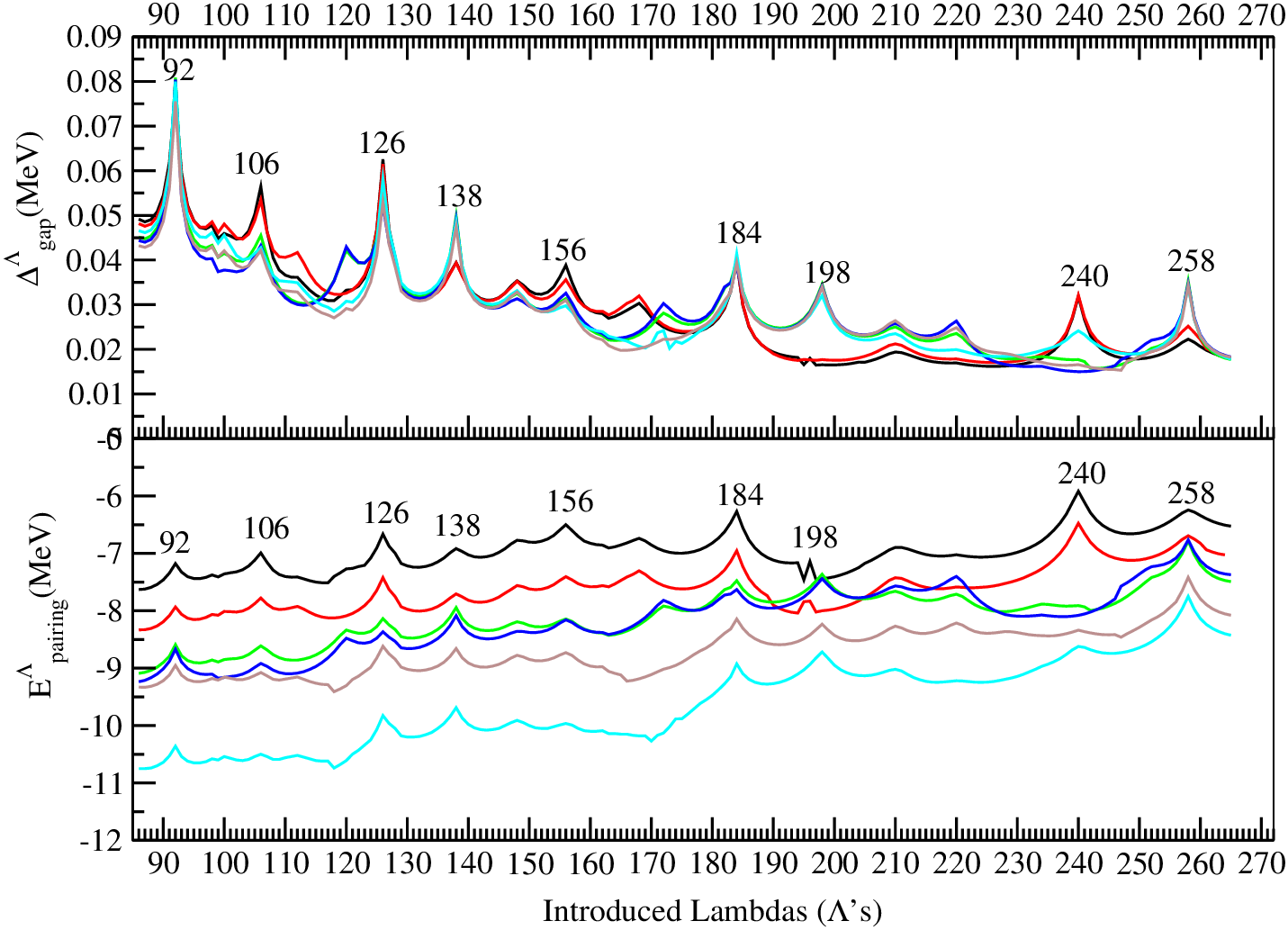}
}
\caption{(color online) The effective $\Lambda$ pairing gap 
($\Delta^\Lambda_{gap}$ ) as a function of $\Lambda$ number as given in upper panel. 
The $\Lambda$ pairing energies ($E^\Lambda_{pairing}$) as a function of $\Lambda$ 
number shown in lower panel.}
\label{pair}
\end{figure}

The outcome of $\Lambda$ pairing energy and effective pairing gaps
for considered multi-$\Lambda$ hypernuclei under investigation 
are plotted in Fig.~\ref{pair}.
Figure~\ref{pair} reveals the minimum contribution of pairing energy 
indicating by peaks at lambda number 92, 106, 126, 138, 156, 184, 198, 240, 258 
which are suggested to be $\Lambda$ shell closure in strangeness sector. 
The suggested magic number on the basis of $\Lambda$ pairing energy and 
effective $\Lambda$ pairing gaps are in close agreement with other signatures of
magicity discussed earlier in the manuscript except $\Lambda =156$ 
which make their appearances in effective $\Lambda$ pairing gap
and pairing energy respectively for the chosen set of multi-$\Lambda$ hypernuclei.

%%%%%%%%%%%%%%%%%%%%%%%%%%%%%%%%%%%%%%%%%%%%%
\subsection{Lambda single-particle energy spectrum}
\begin{figure}
\vspace{0.75cm}
\resizebox{0.50\textwidth}{!}{%
  \includegraphics{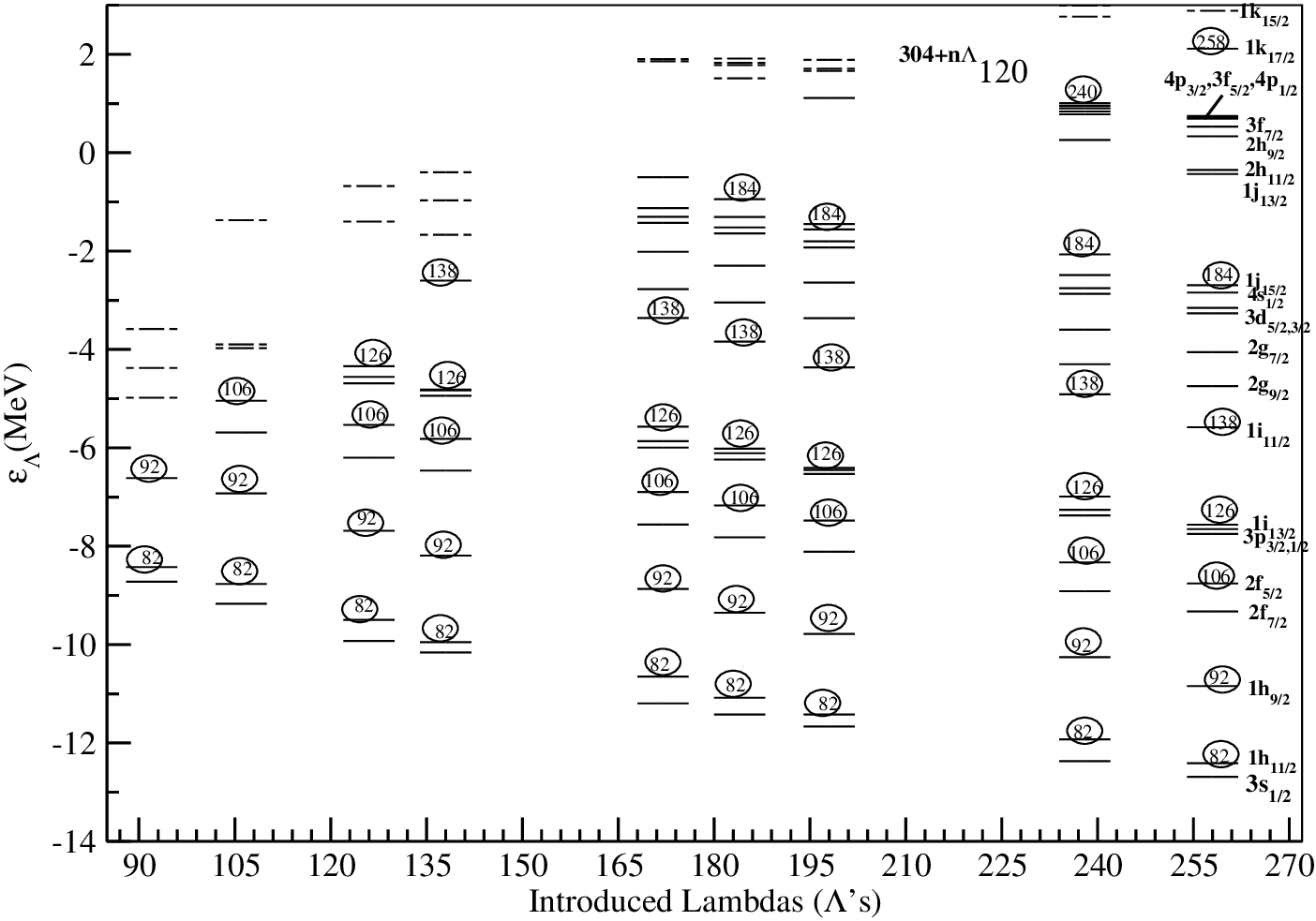}
}
\caption{Lambda single-particle energy levels for $^{304+n{\Lambda}}$120 
multi-$\Lambda$ hypernuclei for $\Lambda$ = 92, 106, 126, 138, 172, 184, 198, 240 
and 258. Filled levels are represented by solid lines whereas dashed 
line represents the unoccupied levels of the given system. }
\label{level120}
\end{figure}

\begin{figure}
\vspace{0.75cm}
\resizebox{0.50\textwidth}{!}{%
  \includegraphics{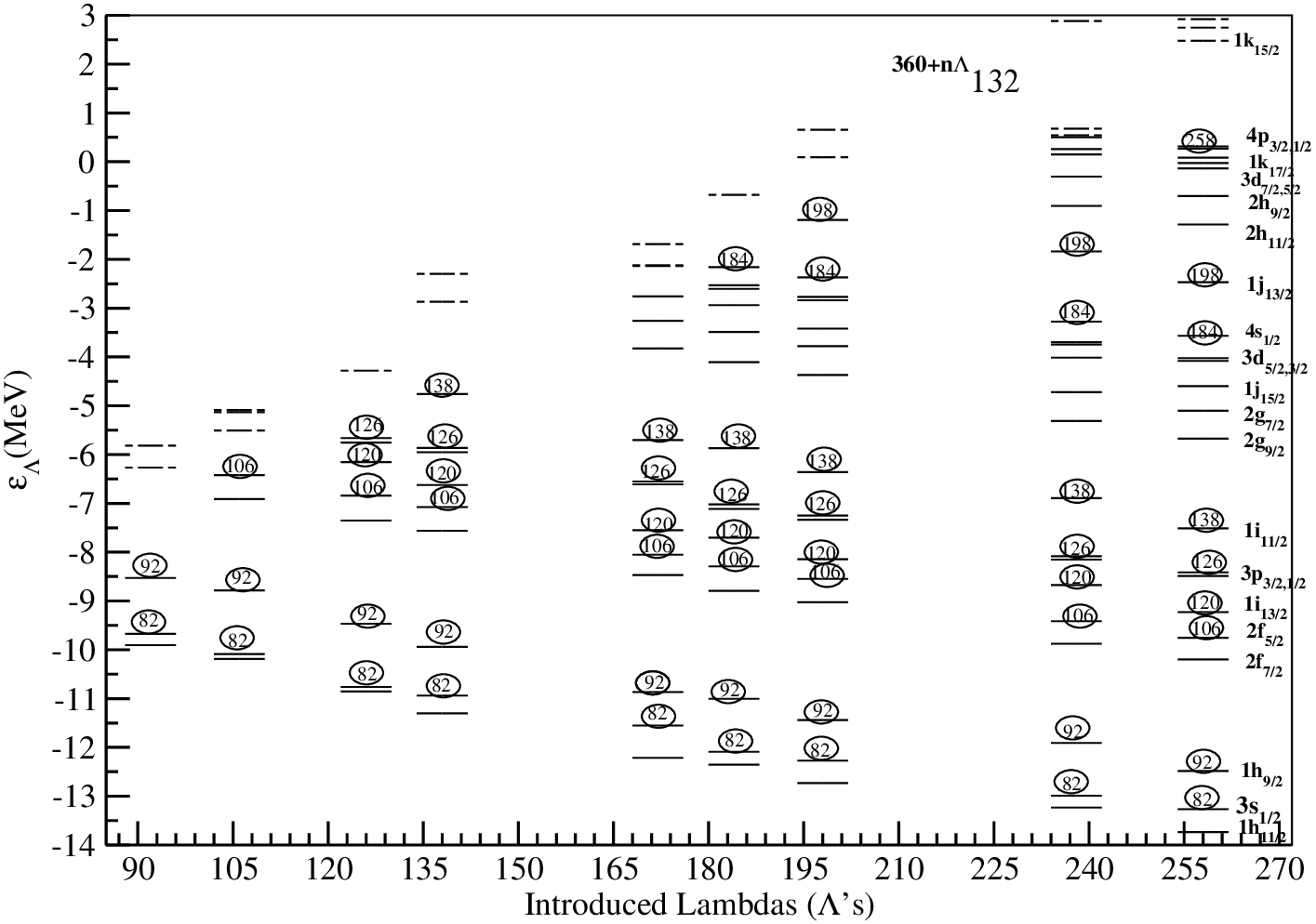}
}
\caption{Same as Fig.~\ref{level120} but for superheavy mass 
$^{360+n{\Lambda}}$132 multi-$\Lambda$ hypernuclei.}
\label{level132}
\end{figure}

\begin{figure}
\vspace{0.75cm}
\resizebox{0.50\textwidth}{!}{%
  \includegraphics{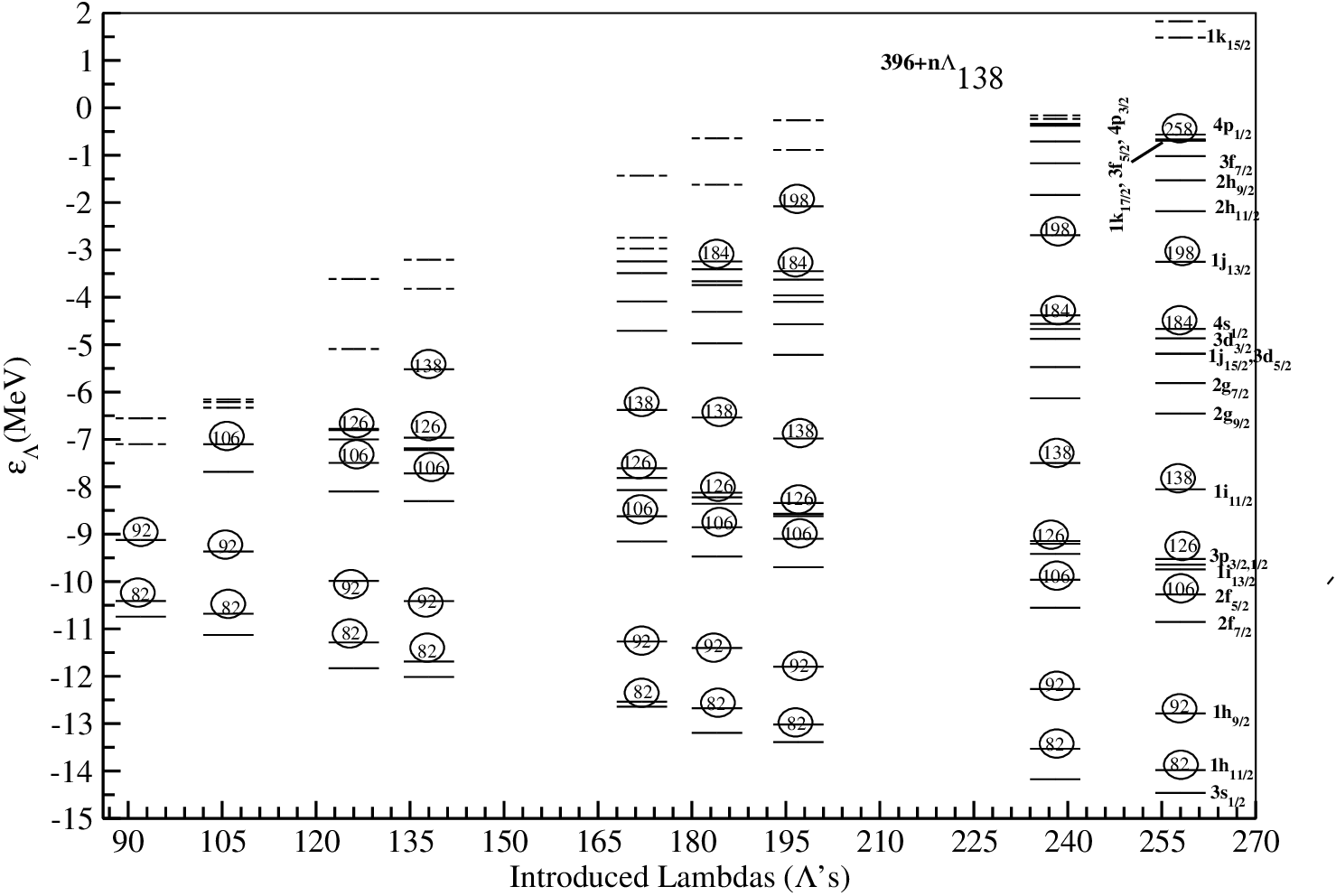}
}
\caption{Same as Fig.~\ref{level120} but for superheavy mass 
$^{396+n{\Lambda}}$138 multi-$\Lambda$ hypernuclei.}
\label{level138}
\end{figure}

The single-particle spectrum reflects any kind of change occurring 
in a system either normal nuclei or hypernuclei.
In order to analyze the effects of hyperon addition to the 
predicted doubly magic nucleonic cores and to confirm or provide
 the support to our calculations, we have plotted the lambda single-particle 
energy levels $^{304+n\Lambda}$120,
 $^{360+n\Lambda}$132 and $^{396+n\Lambda}$138 systems as given 
in Figs.~\ref{level120},\ref{level132} and~\ref{level138}.  
The prescription for filling the lambda energy levels is same as that
of nucleon energy levels filled according to shell model 
scheme with $\Lambda$ spin-orbit potential. However, it is evident
that the single-particle energy gap between lambda levels
is smaller as compared to gaps in nucleon single-particle
energy scheme owing to relatively weaker strength of $\Lambda$
spin-orbit interaction.
In previous article~\cite{ikram2016}, we have analyzed the $\Lambda$ 
single-particle energy levels up to the filling of 82 lambdas 
and predicted the $\Lambda$ magic number from 2 to 82. 
Therefore, in present case we analyze the higher levels above 
the filling of 82 $\Lambda$'s to look for the lambda magicity in higher region.
To make this things, we focused on $^{304+n\Lambda}$120,
 $^{360+n\Lambda}$132 and $^{396+n\Lambda}$138 superheavy 
multi-$\Lambda$ hypernuclei.

After carefully analyzing the single-particle energy diagram
of $^{360+n\Lambda}$ 132 a large gap exists between 1h$_9/2$ and 
2f$_{7/2}$ thus resulting in the appearance of $\Lambda$ = 92 
as strong magic shell closure. This is followed by a relatively
smaller gap between 2$f_{5/2}$ and 1$i_{13/2}$ thus making 
$\Lambda$ = 106 a relatively weak magic number compared to $\Lambda$ = 92. 
Further, the energy gaps between 1$i_{13/2}$ and 3$p_{1/2}$,
 3$p_{1/2}$ and 1$i_{11/2}$ are large compared to the
 previous case. This results in the emergence of relatively
strong magic numbers at $\Lambda$ = 120 and $\Lambda$ = 126, respectively.
 Moreover, the energy gap between 1$i_{11/2}$ and 2$g_{9/2}$ 
is of comparable magnitude to the energy gap corresponding to
 $\Lambda$ = 92 magic number and thus makes $\Lambda$ = 138 a
 strong magic number. In addition, two gaps of almost same magnitude between 4$s_{1/2}$
 to 1$j_{13/2}$ and 1$j_{13/2}$ to 2$h_{11/2}$ which hints 
towards the two strong magic numbers corresponding to $\Lambda$ = 184, 198.
 Moreover, a large energy gap is observed between 4$p_{1/2}$ 
and 1$k_{15/2}$ which results in $\Lambda$ = 258 as strong magic number.
The identical $\Lambda$ magic numbers are reproduced in $^{396+n\Lambda}$138
except for $\Lambda$ = 120 which disappears or gets quenched.
The shell closure at $\Lambda$ = 120 is quenched in $^{304+n\Lambda}$120, however 
large shell gap between 4$p_{1/2}$ to 1$k_{17/2}$ is appeared 
resulting the emergence of $\Lambda$ = 240 magic number.
Some inversion of lambda levels is noticed in 
considered multi-hypernuclear candidates.
For example, 3$s_{1/2}$ filled faster than 1$h_{11/2}$ in case 
of $^{304+n\Lambda}$120, $^{396+n\Lambda}$138 which 
is inverted in case of $^{360+n\Lambda}$132.
The lambda level 4$p_{1/2}$is the last filled level 
in $^{396+n\Lambda}$138 and $^{360+n\Lambda}$132 while this 
level got inverted with 1$k_{17/2}$ in $^{304+n\Lambda}$120. 
Therefore, a inversion  between \{4$p_{1/2}$, 1$k_{17/2}$\} and 
\{3$s_{1/2}$, 1$h_{11/2}$\} is observed.
The inversion or filling of higher levels faster than lower one and hence 
this type of filling is responsible for emerging the variation in 
prediction of new magic number and this is 
due to the spin-orbit interaction.
This is also seen here for example, 120 is appeared to be $\Lambda$ 
magic number in some systems which is quenched in others. 
Also, 240 is appeared only in $^{304+n\Lambda}$120 and 
$^{292+n\Lambda}$120 but missed in other systems.
In general, it is quite worth to mention  that pronounced shell gaps for 
lambda number 92, 106, 126, 138, 184, 198, 240 and 258
are noticed in considered multi-$\Lambda$ hypernuclei and 
being suggest to be strong $\Lambda$ shell closures as well as 
reflected in $\delta_{2\Lambda}$ calculations.

%%%%%%%%%%%%%%%%%%%%%%%%%%%%%%%%
\subsection{Lambda magicity}
The impetus for searching the magic numbers in hypernuclear physics 
earns its motivation from the ever evolving search of magic numbers 
in conventional nuclear physics. 
This encourages us to look for the magicity in strangeness sector. 
We conducted a study on the multi-$\Lambda$ hypernuclei under 
investigation within the powerful framework of spherical relativistic 
mean field with NL3$^{*}$ nucleon-meson parameterization in inclusion 
of $\Lambda$N as well as $\Lambda\Lambda$ interactions. 
Prediction of magic numbers in superheavy region within various 
theoretical models is a debated subject which arouse our interest to 
look for magicity in strangeness sector.
After analyzing the $S_{\Lambda}$ and $S_{2\Lambda}$ for 
the multi-$\Lambda$ hypernuclei, we noticed a 
sudden fall in separation energy is observed 
at ${\Lambda}$ = 92, 106, 126, 138, 184, 198, 240 and 258 and we conclude 
that these numbers might be the possible ${\Lambda}$ magic numbers 
in the multi-$\Lambda$ hypernuclei resulting 
the triply magic system with doubly magic nucleonic cores. 
In order to make a stringent test to our findings and to confirm 
our calculations based on $S_{\Lambda}$ and $S_{2\Lambda}$, we 
made investigation on the basis of two lambda shell gap which 
is considered to be the strong signature of magicity. 
We observed that the predictions made on the basis of $\delta_{2\Lambda}$ 
are in tune with the possibilities on the pretext of 
$S_{\Lambda}$ and $S_{2\Lambda}$. 
A pronounced peak is observed for ${\Lambda}$ = 92, 126, 138, 184, 240 and 258 
in the two lambda shell gap signaling the presence of strong shell closure. 
The peak at $\Lambda$ = 106, 198 are of slightly less magnitude then the others.
Moreover, in favour of $\Lambda$ predictions the investigation of $\Lambda$ 
pairing energy and effective $\Lambda$ pairing gap has been made and its outcome 
are found to be in very close agreement with the survey made on the pretext 
of separation energy and shell gaps.
Further, in order to support in these calculation and to reveal the magic 
nature of these numbers, we analyzed the lambda single-particle energy 
spectrum of $^{304+n\Lambda}$120, $^{360+n\Lambda}$132 
and $^{396+n\Lambda}$138. 
Inspection of the single-particle energies for these 
chosen multi-$\Lambda$ hypernuclei reveals a large shell 
gaps at ${\Lambda}$ = 92, 106, 120, 126, 138, 184, 198, 240 and 258. 
The shell gaps at ${\Lambda}$ = 120 does not appear in all investigated 
candidates and make its appearance only in $^{360+n\Lambda}$132. 
This leads us to the conclusion that ${\Lambda}$ = 120 is not a strong 
magic number in strangeness physics and might be considered as very 
feeble magic number, however, it is predicted to be a strong proton 
magic number supported by many theoretical investigations. 
The predicted $\Lambda$ magic number in present calculations quite 
resembles with the nuclear magic number of superheavy region.
Therefore, it can be concluded that ${\Lambda}$ magicity resembles 
quite closely with the predicted magic numbers in 
enigmatic superheavy valley. 
%The present predicted magic numbers are in consonance 
%with the theoretical estimations in the Refs.~\cite{ccc}. 
The investigations about magic numbers might serve as a 
significant input to reveal important 
information regarding new nucleonic shell closures.

\begin{table}
\caption{Lambda magic number produced in various considered 
multi-$\Lambda$ hypernuclei are tabulated here.}
\vskip0.5cm
\renewcommand{\tabcolsep}{0.13cm}
%\renewcommand{\arraystretch}{1.2}
%\begin{ruledtabular}
{
\footnotesize

\begin{tabular}{cccccccccc}
\hline\hline
\multicolumn{1}{c}{Hypernuclei}&\multicolumn{9}{c}{Lambda magic numbers}\\
\cline{1-1}\cline{2-10}
%Hypernuclei&&&&&&&&&&&&&\\
\hline
$^{292+n\Lambda}$120 &92&106&-  &126&138&184&198&240&258\\
$^{304+n\Lambda}$120 &92&106&-  &126&138&184&198&240&258\\
$^{360+n\Lambda}$132 &92&106&120&126&138&184&198&-  &258\\
$^{370+n\Lambda}$132 &92&106&-  &126&138&184&198&-  &258\\
$^{336+n\Lambda}$138 &92&106&-  &126&138&184&198&-  &258\\
$^{396+n\Lambda}$138 &92&106&-  &126&138&184&198&-  &258\\
\hline\hline
\end{tabular}
}
%\end{ruledtabular}
\label{tab2}
\end{table}

%%%%%%%%%%%%%%%%%%%%%%%%%%%%%%%%%%%%%%%%%%%%%%%%%%%%%%%%%%%%%%%
\subsection{Lambda and nucleon spin-orbit interaction potentials}
\begin{figure}
\vspace{0.75cm}
\resizebox{0.49\textwidth}{!}{%
  \includegraphics{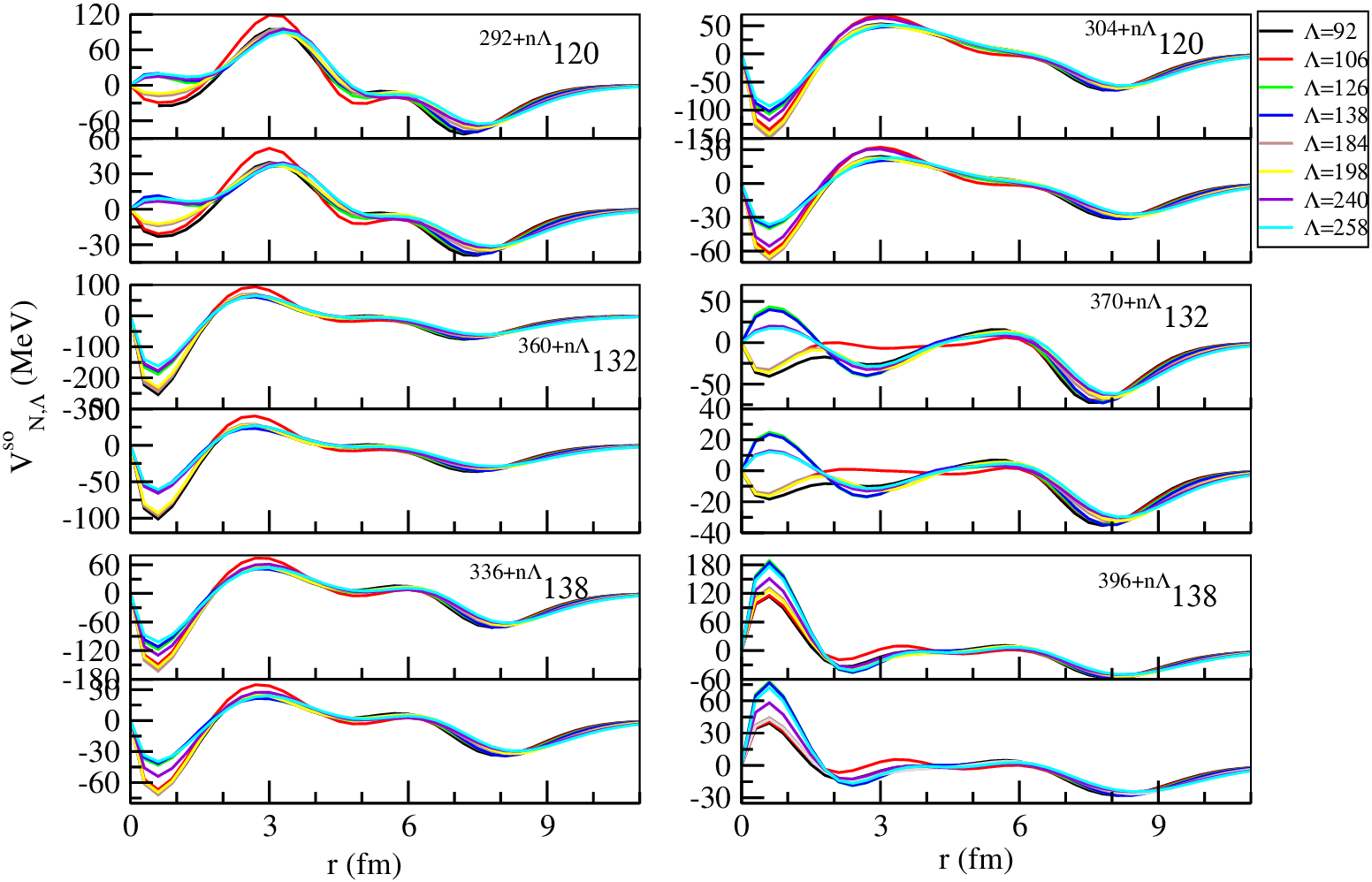}
}
\caption{(color online) Spin-orbit interaction potentials of 
nucleons and $\Lambda$ hyperons for considered superheavy 
multi-$\Lambda$ hypernuclei for lambda number 92, 106, 126, 
138, 184, 198, 240 and 258 as function of radial parameter. 
The upper part in each panel represent the nucleon spin-orbit and 
the lower one representing the $\Lambda$ spin-orbit potentials.}
\label{sorbit}
\end{figure}

It has already been mentioned that YN data available is very scanty.
 Moreover, a very little know about the spin-dependent interaction
 such as spin-orbit force and hence the spin-orbit splitting in
 $\Lambda$N interaction has become a subject of study in strangeness physics.
 Lambda spin-orbit coupling has been found to be smaller than that
 of nucleons and no definite value is produced by the 
experiment~\cite{bruck1976,may1981,may1983,ajimura2001}.
 The smallness of $\Lambda$ spin-orbit splitting in $\Lambda$N
 interaction was first suggested in the (K$^-$, $\pi^-$)
 reaction on $^{16}O$~\cite{bruck1976}. The smaller magnitude of 
$\Lambda$ spin-orbit potential is encountered due to the 
presence of antisymmetric part of spin-orbit force in $\Lambda$N 
interaction and also by the energy difference
 of single-particle states of j = l - 1/2 and j = l + 1/2.
 The spin-orbit force in $\Lambda$N interaction has
 an antisymmetric part \{l.($S_\Lambda-S_N$)\} in addition
 to the usual force.  These two parts of spin-dependent 
force tend to cancel in $\Lambda$N interaction which leads
 to a small amount of lambda spin-orbit splitting in $\Lambda$ 
single particle states.
 However, for quantitative description of spin-orbit force,
 various theoretical relativistic and non-relativistic 
 models with suitable effective force have been employed.
 In non-relativistic model, the spin-orbit force is 
included manually while it is naturally developed in relativistic approaches 
such as RMF and spin-orbit potential emerges
 automatically with the empirical strength.
 This coupling is very crucial for quantitative
 description of nuclear or hypernuclear structure
 as well as in order to reproduce the empirical magic numbers.
 In present case, we analyzed the nucleon as well as
 $\Lambda$ spin-orbit interaction potentials for considered
 multi-$\Lambda$ hypernuclei as shown in Fig. ~\ref{sorbit},
 where we found a consistency 
in nucleonic and hyperonic splitting as previous
 predictions that implies reduction of $\Lambda$
 spin-orbit is about 30 - 40\% of nucleonic case.
 For example; for $^{292+n\Lambda}$120 at r = 3 fm the 
$V^{so}_N$ is about of 120 MeV and this value reduced
 to 45 MeV for $V^{so}_\Lambda$ that comes out to be around
 40\% of nucleonic potentials. 
It is quite evident that both the potentials 
are different by their depth but have the simar interaction profile.
Further inspection reveal that injection of $\Lambda$'s 
significantly affects the nucleon as well as $\Lambda$ 
spin-orbit potential to a large extent. 

\begin{figure}
\vspace{0.75cm}
\resizebox{0.50\textwidth}{!}{%
  \includegraphics{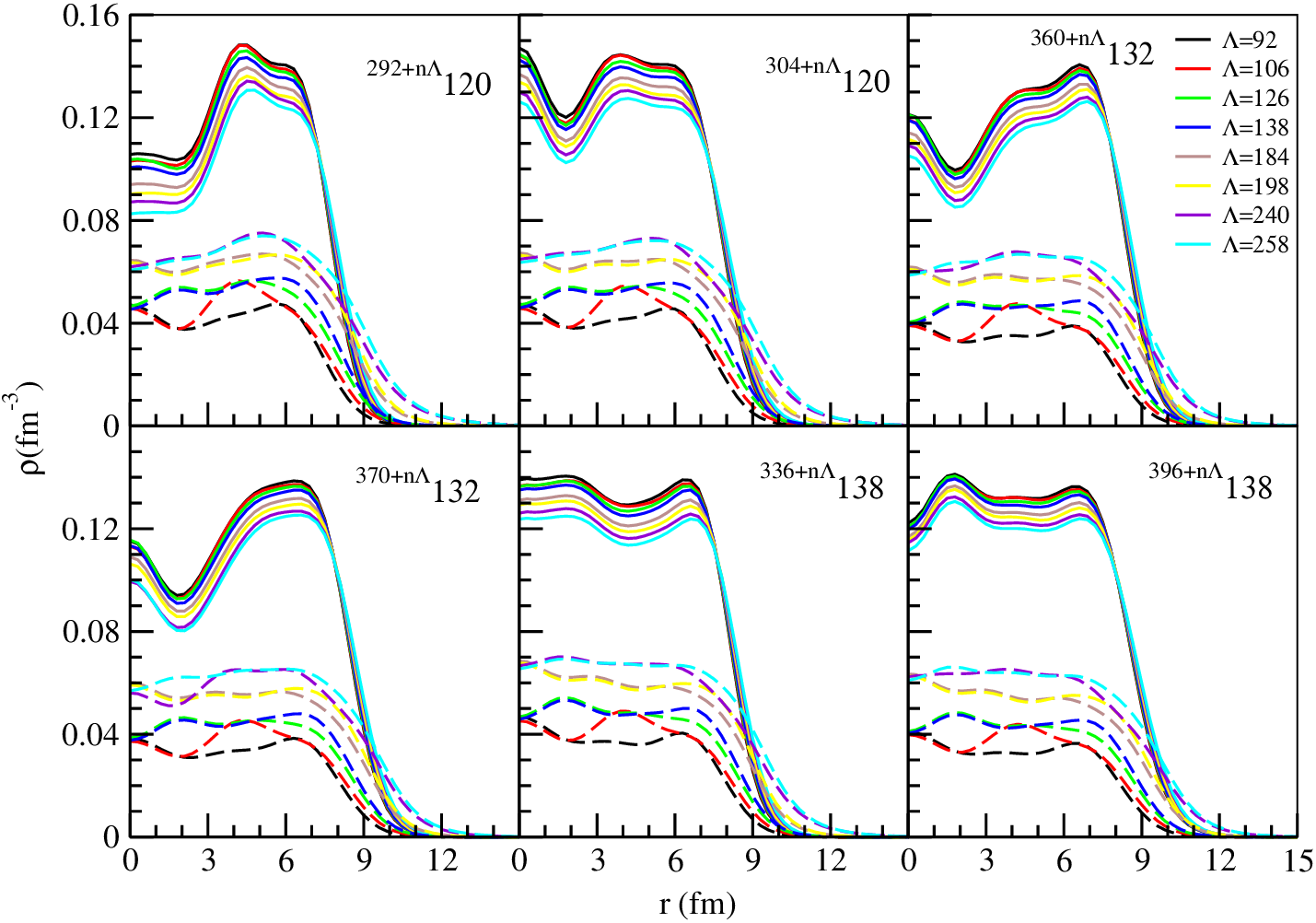}
}
\caption{(color online) Nucleon and $\Lambda$ density profile 
for considered superheavy multi-$\Lambda$ hypernuclei for 
$\Lambda$ number 92, 106, 126, 138, 184, 198, 240 and 258. 
Nucleon density is represented by solid lines whereas  
$\Lambda$ density by dashed lines.}
\label{den}
\end{figure}

\begin{figure}
\vspace{0.75cm}
\resizebox{0.50\textwidth}{!}{%
  \includegraphics{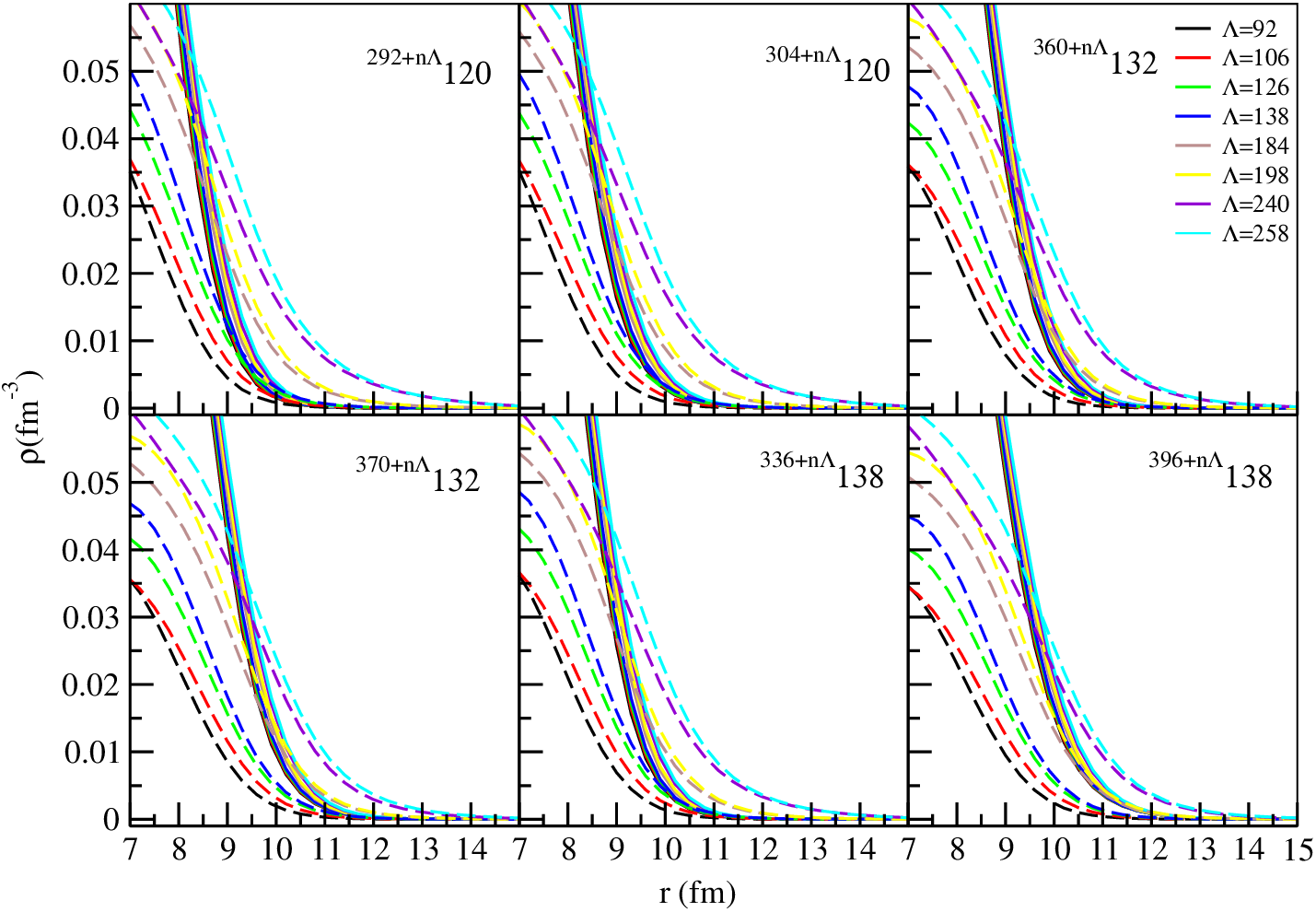}
}
\caption{(color online) Same as Fig.~\ref{den} but for peripheral region}.
\label{pheri}
\end{figure}

%%%%%%%%%%%%%%%%%%%%%%%%%%%%%%55
\subsection{Density profile}
It is worthy to mention that the injected $\Lambda$ hyperon affects the
every piece of physical observables such as size, shape, 
binding energy, density distribution etc. 
Therefore, it is quite interesting as well as important to study 
the effects of large number of $\Lambda$ on the nuclear density 
and hence the distribution of nucleons and $\Lambda$ hyperons in considered 
multi-$\Lambda$ systems is analyzed by means of density profile 
as shown in Figure~\ref{den}. 
From this figure, we noticed that the successive addition of 
$\Lambda$ hyperons reduces the magnitude of central nuclear density 
resulting increases the matter radii of the system. 
This is happened because injected lambdas are sitting in higher orbitals 
that attributed to increase the size of the system.
For example; in case of $^{292+n\Lambda}$120, the magnitude of 
central density for $\Lambda$ = 92 is 0.11 $fm^-3$ whereas 
this value reaches to 0.08 $fm^-3$ for $\Lambda$ = 258 
indicating a reduction of central density. 
In other words, we can say that the excessive addition  of $\Lambda$ 
hyperons depressed the central region of nuclear core. 
This implies that the presence of large number of $\Lambda$'s pushes 
out the nucleons from the central region. 
Moreover, only magnitude of central density is reduced 
while the shape of profile is unaltered. 
Further inspection reveals that the 
lambda density increases with successive addition of 
$\Lambda$ hyperons and that is obvious. 
Since, there is no change of nucleon numbers and hence 
no anomalous behaviour of nuclear density is encountered 
in considered multi-$\Lambda$ hypernuclei. 
Moreover for the sake of skin/halo structure, we 
analyzed the behaviour of lambda and nucleon 
density profile at peripheral region. 
For this we make a plot of density profile at radial parameter r = 7 fm 
to 15 fm shown in Fig.~\ref{pheri}. 
The nucleon density dies at r = 11 fm while $\Lambda$ density 
extended to 14 fm as evident in case of $^{292+n\Lambda}$120 
and $^{304+n\Lambda}$120 and others also. 
This dictate that the presence of large number of $\Lambda$'s 
pushes the lambda itself towards the periphery of the system 
and this kind of behaviour of $\Lambda$ distribution form 
a $\Lambda$ skin structure.

%%%%%%%%%%%%%%%%%%%%%%%%%%%%%%%%
\section{Summary and conclusions}
In present work, we performed spherical relativistic calculations 
with $\Lambda$N and $\Lambda\Lambda$ interactions in multi-$\Lambda$ 
hypernuclei and postulated the possible $\Lambda$ magic
number i.e., 92, 106, 120, 126, 138, 184, 198, 240, 258.
These predictions have been made by the survey employing various 
signatures of magicity in strangeness sector like one- and two-$\Lambda$ 
separation energy and two lambda shell gaps.
We witnessed prominent peaks as well as single-particle energy gaps
for lambda numbers 92, 126, 184 and 258. 
Further, the predictions made in multi-$\Lambda$ hypernuclei under 
study resembles quite closely with the magic numbers in 
conventional nuclear theory suggested by various relativistic and 
non-relativistic theoretical models and these $\Lambda$ predictions 
also support the confirmation of nucleon magicity of superheavy regime.
%These lambda predictions also give a one more strong evidence about 
%the confirmation of nuclear magicity of superheavy regime.
Thus, it can be inferred that YN interaction is of similar nature
as NN interaction except its weaker strength compared to nucleonic case. 
The appearance of new lambda shell closures other than 
nucleonic one those predicted by various theoretical approaches 
in superheavy mass regime can be attributed to the relatively weak 
strength of spin-orbit interaction in strange sector. 
Further, to lend a support to our calculations, we explore the 
occurrence of spherical shell closures by doing survey on the 
basis of lambda pairing energy and effective lambda pairing gap 
and the results are quite consistent with the predictions made 
using $S_{\Lambda}$, $S_{2\Lambda}$ and $\delta_{2\Lambda}$ except 
for the appearance of magic number corresponding to $\Lambda$ = 156 
which emerges in lambda effective pairing gap and pairing energy 
within the considered multi-$\Lambda$ hypernuclei.
Lambda single-particle spectrum is also analyzed to mark the 
energy shell gap for further strengthening the predictions made 
on the basis of separation energy and shell gaps.
Lambda and nucleon spin-orbit interactions are analyzed to confirm the 
reduction in magnitude of $\Lambda$ spin-orbit interaction to the 
nucleonic case, however interaction profile is similar in both the cases.
It is also concluded that addition of $\Lambda$'s significantly 
affects the nucleon as well as $\Lambda$ spin-orbit potential. 
Lambda and nucleon density distributions have been made to 
further confirm the impurity effect of $\Lambda$ hyperons 
which make the reduction in magnitude of central density of the cores.
Lambda skin structure is also reported.
The stability attributed to the doubly magic superheavy cores by addition of
hyperons may serve as an impetus to the experimentalists
for synthesizing these triply magic system in near future, 
might provide significant input and theirby may shed some 
light pertaining to novel features of new nucleonic shell closures. 
Further, superheavy nucleonic cores (doubly magic)
are seen to have more affinity to absorb large number
of hyperons. Hence, it can be said that such systems 
replicate the strange hadronic matter containing multiple 
degrees of strangeness (multi-strange) such as $\Sigma$'s, $\Xi$'s 
and hyperons which has got huge importance in nuclear 
astrophysics like neutron stars, hyperon stars etc.

\section{Acknowledgments}
One of the authors (MI) would like to acknowledge the 
hospitality provided by Institute of Physics (IOP), 
Bhubaneswar where the parts of this work was carried out.

%%%%%%%%%%%%%%%%%%%%%%%%%%


\begin{thebibliography}{99}
\bibitem{NRS77} M. M.  Nagels,  T. A. Rijken, and J. J. de Swart, Phys. Rev. D {\bf15}, 2547(1977).
\bibitem{NRS79} M. M. Nagels, T. A. Rijken, and J. J. de Swart, Phys. Rev. D {\bf20}, 1633(1979).
\bibitem{MRS89} P. M. M. Maessen, T. A. Rijken, and J. J. de Swart, Phys. Rev. C {\bf40}, 2226(1989).
\bibitem{RSY99} T. A. Rijken, V. J. G. Stoks, and Y. Yamamoto, Phys. Rev. C {\bf59}, 21(1999).
\bibitem{RY06a} T. A. Rijken, and Y. Yamamoto, Phys. Rev. C {\bf73}, 044008(2006).
\bibitem{NRY15b} M. M. Nagels, T. A. Rijken, and Y. Yamamoto, arXiv:1501.06636, (2015).
\bibitem{NRY15a} M. M. Nagels, T. A. Rijken, and Y. Yamamoto, arXiv:1504.02634, (2015).
\bibitem{HHS89} B. Holzenkamp, K. Holinde, and J. Speth, Nucl. Phys. A {\bf500}, 485(1989).
\bibitem{RHS94} A. Reuber, K. Holinde, and J. Speth, Nucl. Phys. A {\bf570}, 543(1994).
\bibitem{HM05} J. Haidenbauer, and U.-G. Meißner, Phys. Rev. C {\bf72}, 044005(2005).
\bibitem{PHM06} H. Polinder, J. Haidenbauer, and U.-G. Meißner, Nucl. Phys. A {\bf779}, 244(2006).
\bibitem{HPKMNW13} J. Haidenbauer, S. Petschauer, N. Kaiser, U.-G. Meißner, A. Nogga, and W. Weise, Nucl. Phys. A {\bf915}, 24(2013).
\bibitem{H13} J. Haidenbauer, Nucl. Phys. A {\bf914}, 220(2013).
\bibitem{FSN07} Y. Fujiwara, Y. Suzuki, and C. Nakamoto, Prog. Part. Nucl. Phys. {\bf58}, 439(2007).
\bibitem{KF09} M. Kohno, and Y. Fujiwara, Phys. Rev. C {\bf79}, 054318(2009).
\bibitem{M49} M.G. Mayer, Phys. Rev. {\bf75}, 1969(1949).
\bibitem{HJS49} O. Haxel , H. H. D. Jensen,  and H. E. Suess, Phys. Rev. {\bf75}, 1766(1949).
%\bibitem{BBFH57} E. M. Burbidge , G. R. Burbidge , W. A. Fowler and F. Hoyle, Rev. Mod. Phys. {\bf29}, 547(1957).
\bibitem{BM69} A. Bohr and B. Mottelson, Nuclear Structure (New York: Benjamin) (1969).
\bibitem{PS80} P. Ring and P. Schuck The Nuclear Many-Body Problem (Berlin: Springer) (1980).
\bibitem{H99} K. Heyde, Basic Ideas and Concepts in Nuclear Physics(Bristol: Institute of Physics Publishing) (1999).
\bibitem{W74} J. D. Walecka, Ann. Phys., NY {\bf83}, 491(1974).
\bibitem{SW86} B. D. Serot and J. D. Walecka, Adv. Nucl. Phys. {\bf16}, 1(1986).
\bibitem{R89} P. G. Reinhard, Rep. Prog. Phys. {\bf52}, 439(1989).
\bibitem{HTWTC91} D. Hirata, H. Toki, T. Watabe, I. Tanihata and B. V. Carlson Phys. Rev. C {\bf44}, 1467(1991).
\bibitem{R96} P. Ring, Prog. Part. Nucl. Phys. {\bf37} 193(1996).
\bibitem{BGH85} M. Brack, C. Guet, and B. Hakansson, Phys. Rep. {\bf123}, 275(1985).
\bibitem{S94} A. Sobiczewski Fiz. Elem. Chastits At. Yadra {\bf295}, 25(1994).
\bibitem{MNMS95} P. M\"oller, J. R. Nix, W. D. Myers and W. J. Swiatecki, At. Data Nucl. Data Tables {\bf59}, 185(1995).
\bibitem{MNK97} P. M\"oller, J. R. Nix,and K. Kratz, At. Data Nucl. Tables {\bf66}, 131(1997).
\bibitem{D05} V. Y. Denisov Phys. At. Nucl. {\bf68}, 1133(2005).
\bibitem{BBGD01} J. F. Berger, L. Bitaud, M. Girod and K. Dietrich, Nucl. Phys. A {\bf685}, 644(2001).
\bibitem{BNR01} M. Bender, W. Nazarewicz, and P.-G. Reinhard, Phys. Lett.
B, {\bf515}, 42(2001).
\bibitem{BRRMG99} M. Bender, K. Rutz, P.-G. Reinhard, J. A. Maruhn, and W.
Greiner. Phys. Rev. C, {\bf60}, 034304(1999).
\bibitem{BRRMG98} T. B\"urvenich, K. Rutz, M. Bender, P. G. Reinhard, J. A. Maruhn, and W. Greiner, Eur. Phys.J. A {\bf3}, 139(1998). 
\bibitem{KBNRVC00} A. T. Kruppa, M. Bender, W. Nazarewicz, P.G. Reinhard, T. Vertse and S. \"Cwiok, Phys. Rev. C {\bf61}, 034313(2000).
\bibitem{LMZ00} S. Liran, A. Marinov and N. Zeldes, Phys. Rev. C {\bf62}, 047301(2000).
\bibitem{LMZ000} S. Liran, A. Marinov and N. Zeldes, Phys. Rev. C {\bf66}, 024303(2000). 
%\bibitem{} \bibitem{} \bibitem{MG68} U. Mosel, and W. Greiner, Z. Phys. {\bf 217} 256, 1968.
%\bibitem{NTSWGMS69} S. G. Nilsson, C. F. Tsang, S. Szyman\'ski, S. Wycech, 
%C. Gustafson,I. Lamm, P. M\''oller, and A. N. B. Sobiczewski, Nucl. Phys. A {\bf 131},1 (1969).
%\bibitem{YR08} Y. Yamamoto, and T. A. Rijken, Nucl. Phys. A {\bf 804}, 139(2008).
\bibitem{LSRG96} G. Lalazissis, M. Sharma, P. Ring, and Y. Gambhir, Nucl.
Phys. A, {\bf608}, 202(1996)
\bibitem{CDMNH96} S. \'Cwiok, J. F. Dobaczewski, P. Magierski  and W. Nazarewicz 
 P. H. Heenen, Nucl. Phys. A, {\bf611}, 211(1996).
\bibitem{RBBSRMG96}  K. Rutz, M. Bender, T. B\"urvenich, T. Schilling, P. G. Reinhard,
J. A. Maruhn, W. Greiner, Rev. C, {\bf56}, 238(1997).
\bibitem{BRB98} T. B\"urvenich, K. Rutz, M. Bender et al, Eur. Phys. J, A {\bf3},
139(1998).
\bibitem{KBN00} A. T. Kruppa, M. Bender, W. Nazarewicz et al. Phys. Rev. C,
{\bf61}, 034313(2000).
\bibitem{SPSCV04} T. Sil, S. K. Patra, B. K. Sharma, M. Centelles, and X. Vinas,
Phys. Rev. C{\bf69}, 044315(2004).
\bibitem{mbhuyan} M. Bhuyan and S. K. Patra, Mod. Phys. Lett. A {\bf27}, 1250173 (2012).
\bibitem{ZMZGT05} W. Zhang, J. Meng, S. Q. Zhang, L. S. Geng, and H. Toki,
Nucl. Phys. A, {\bf753}, 106(2005).
\bibitem{NS14} H. Nakada and K. Sugiura, Prog. Theor. Exp. Phys. {\bf2}, 033D02 (2014).
\bibitem{N13} H. Nakada, Phys. Rev. C {\bf87}, 014336(2013).
\bibitem{BGG91} J. F. Berger, M. Girod and D. Gogny, Comput. Phys. Commun.{\bf63}, 365(1991).
\bibitem{GHGP09} S. Goriely, S. Hilaire, M. Girod and S. P\"eru, Phys. Rev. Lett. {\bf102}, 242501(2009).
\bibitem{IEAA16} M. Ismail, A. Y. Ellithi, A. Adel, and Hisham Anwer, J. Phys.G: Nucl. Part. Phys. {\bf43}, 015101(2016).
\bibitem{ikram2016} M. Ikram, Asloob A. Rather, A. A. Usmani, B. Kumar and S. K. Patra Int. J. Mod. Phys. E {\bf25}, 1650103(2016).
\bibitem{sugahara1994}
Y. Sugahara and H. Toki, Prog. Theor. Phys. {\bf92}, 803(1994).
\bibitem{vretenar1998}
D. Vretenar, W. Po\"schl, G. A. Lalazissis and P. Ring, Phys. Rev. C {\bf57}, R1060(1998).
\bibitem{lu2003}
H. -F. L\"u, J. Meng, S. Q. Zhang and S. G. Zhou, Eur. Phys. J. A. {\bf17}, 19(2003).
\bibitem{win2008}
M. T. Win and K. Hagino, Phys. Rev. C {\bf78}, 054311(2008).
\bibitem{ikram14}
M. Ikram, S. K. Singh, A. A. Usmani and S. K. Patra, Int. J. Mod. Phys. E {\bf23}, 1450052(2014).
\bibitem{glendenning1993}
N. K. Glendenning, D. Von-Eiff, M. Haft, H. Lenske and M. K. Weigel, Phys. Rev. C {\bf48}, 889(1993).
\bibitem{mares1994}
J. Mares and B. K. Jennings, Phys. Rev. C {\bf49}, 2472(1994).
\bibitem{rufa1990}
M. Rufa, J. Schaffner, J. Maruhn, H. St\"ocker and W. Greiner, Phys. Rev. C {\bf42}, 2469(1990).
\bibitem{schaffner2002}
J. Schaffner, M. Hanauske, H. St\"ocker and W. Greiner, Phys. Rev. Lett. {\bf89}, 171101(2002).
\bibitem{shen2006}
H. Shen, F. Yang and H. Toki, Prog. Theor. Phys. {\bf115}, 325(2006).
\bibitem{schaffner1994} 
J. Schaffner, C. B. Dover, A. Gal, C. Greiner, D. J.
Millener and H. St\"ocker, Ann. Phys. (N.Y.) {\bf235}, 35(1994).
\bibitem{schaffner1993}
J. Schaffner, C. B. Dover, A. Gal, C. Greiner and H. St\"ocker, Phys. 
Rev. Lett. {\bf71}, 1328(1993).
\bibitem{lalazissis09}
G. A. Lalazissis, S. Karatzikos, R. Fossion, D. Pena Arteaga, A. V. Afanasjev and P. Ring, Phys. Lett. B {\bf671}, 36(2009). 
\bibitem{dover1984}
C. B. Dover and A. Gal, Prog. Part. Nucl. Phys. {\bf12}, 171(1984).
\bibitem{chiapparini09}
M. Chiapparini, M. E. Bracco, A. Delfino, M. Malheiro, D. P. Menezes, C. Providencia, Nucl. Phys. A {\bf826}, 178(2009).
\bibitem{GM91}
N. K. Glendenning, S. A. Moszkowski, Phys. Rev. Lett. {\bf67}, 2414(1991).
\bibitem{BMP58} A. Bohr, B. R. Mottelson, and D. Pines, Phys. Rev. C {\bf110}, 936(1958)
\bibitem{RO84} P.G. Reinhard,and E. W. Otten, Nucl. Phys. A {\bf420}, 173(1984).
\bibitem{BHR03} M. Bender, P. H. Heenen, and P. G. Reinhard, Rev. Mod. Phys. {\bf75}, 121(2003).
\bibitem{BF76} J. Blocki, M. Flocard, Nucl. Phys. A, {\bf273}, 45(1976).
\bibitem{bruck1976} W. Br\"uckner et. al., Phys. Lett. B {\bf79}, 157 (1978).
\bibitem{may1981} M. May et. al., Phys. Rev. Lett. {\bf47}, 1106 (1981).
\bibitem{may1983} M. May et. al., Phys. Rev. Lett. {\bf51}, 2085 (1983).
\bibitem{ajimura2001} S. Ajimura et. al., Phys. Rev. Lett. {\bf86}, 4255 (2001).


\end{thebibliography}
\end{document}